\documentclass[a4paper,11pt]{article}
\usepackage{aaskaiid}
\usepackage{units}
\usepackage{xspace}
\usepackage{comment}
\usepackage{graphicx}
\usepackage{subcaption}
\usepackage{orcidlink}

\newcommand{\Xmax}{$X_{\rm max}$\xspace}
\newcommand{\Xmaxmath}{X_{\rm max}}

\newcommand{\Xrit}{$X_{\rm RIT}$\xspace}

\title{Origins of Cosmic Rays in the Galactic-extragalactic Transition Energy Range}
\ShortTitle{Origins of high-energy cosmic rays}

\author*[a,b]{Arthur~Corstanje\orcidlink{0000-0001-5992-6228}} 
\author[c, x]{Subhadip~Saha\orcidlink{0000-0003-2435-8317}}

\author[d]{Sjoerd~Bouma\orcidlink{0000-0002-6959-2302}}
\author[e]{Justin~Bray\orcidlink{0000-0002-0963-0223}}
\author[a,b]{Stijn~Buitink\orcidlink{0000-0002-6177-497X}}
\author[a]{Vital~De Henau\orcidlink{0009-0003-0337-3558}}
\author[f]{Edwin~Dickinson\orcidlink{0000-0003-0834-4708}}
\author[j,k]{Brian~Hare\orcidlink{0000-0001-5138-1235}}
\author[c]{Andreas~Haungs\orcidlink{0000-0002-9638-7574}}
\author[l,v]{Haoning~He\orcidlink{0000-0002-8941-9603}}
\author[b,a,m]{J\"org~H\"orandel\orcidlink{0000-0001-6604-547X}}
\author[c,a]{Tim~Huege\orcidlink{0000-0002-2783-4772}}
\author[f]{Clancy~James\orcidlink{0000-0002-6437-6176}}
\author[d]{Philipp~Laub\orcidlink{0009-0003-2617-9109}}
\author[l]{Xingyu Li\orcidlink{0009-0003-1209-2643}}
\author[c]{Hermann-Josef~Mathes\orcidlink{0000-0002-0680-040X}}
\author[b,m]{Katharine~Mulrey\orcidlink{0000-0001-8026-8020}}
\author[d,n]{Anna~Nelles\orcidlink{0000-0002-1720-6350}}
\author[w]{Felix~Schl\"uter\orcidlink{0000-0002-5545-4363}}

\author[j]{Olaf~Scholten\orcidlink{0000-0003-3649-1254}}
\author[e]{Ralph~Spencer\orcidlink{0009-0009-6015-1787}}
\author[k]{Christopher~Sterpka\orcidlink{0000-0001-8217-0836}}
\author[d]{Karen~Terveer\orcidlink{0009-0002-9594-0419}}
\author[p]{Satyendra~Thoudam\orcidlink{0000-0002-7066-3614}}
\author[q]{Gia~Trinh\orcidlink{0000-0002-5352-5092}}
\author[k]{Paulina~Turekova\orcidlink{0009-0006-1262-7507}}
\author[c]{Darko~Veberic\orcidlink{0000-0003-2683-1526}}
\author[c]{Keito~Watanabe\orcidlink{0000-0003-0599-4035}}

\author[s,t]{Chao~Zhang\orcidlink{0000-0001-9366-0056}}
\author[u]{Pengfei~Zhang\orcidlink{0000-0002-6855-5315}}
\author[l]{Yi~Zhang\orcidlink{0000-0001-6223-4724}}

\ShortName{Arthur Corstanje et al.} 

\newcommand{\affilASTRON}{Netherlands Institute for Radio Astronomy (ASTRON), Dwingeloo, The Netherlands}
\newcommand{\affilCanTho}{Physics Education Department, School of Education, Can Tho University, Campus~II, 3/2 Street, Ninh Kieu District, Can Tho City, Viet Nam}
\newcommand{\affilCurtin}{International Centre for Radio Astronomy Research, Curtin University, Bentley, 6102, WA, Australia}
\newcommand{\affilDESY}{Deutsches Elektronen-Synchrotron DESY, Platanenallee~6, 15738 Zeuthen, Germany}
\newcommand{\affilErlangen}{Erlangen Centre for Astroparticle Physics, Friedrich-Alexander-Universit\"at Erlangen-N\"urnberg, 91058 Erlangen, Germany}
\newcommand{\affilGorlitz}{Deutsches Zentrum f\"ur Astrophysik, Postplatz~1, 02826 Görlitz, Germany}
\newcommand{\affilGroningen}{Kapteyn Astronomical Institute, University of Groningen, P.O.~Box 72, 9700 AB Groningen, Netherlands}
\newcommand{\affilHefei}{School of Astronomy and Space Science, University of Science and Technology of China, Hefei 230026, China}
\newcommand{\affilKanpur}{Department of Physics, Indian Institute of Technology Kanpur, Kanpur, UP-208016, India}
\newcommand{\affilKeyNanjing}{Key Laboratory of Modern Astronomy and Astrophysics, Nanjing University, Ministry of Education, Nanjing 210023, China}
\newcommand{\affilKIT}{Institut f\"ur Astroteilchenphysik, Karlsruhe Institute of Technology (KIT), P.O.~Box 3640, 76021 Karlsruhe, Germany}
\newcommand{\affilKhalifa}{Department of Physics, Khalifa University, P.O.~Box 127788, Abu Dhabi, United Arab Emirates}
\newcommand{\affilManchester}{Jodrell Bank Centre for Astrophysics, Department of Physics and Astronomy, University of Manchester, Manchester M13 9PL, UK}
\newcommand{\affilMaxPlanck}{Max-Planck Institut f\"ur Astrophysik, Karl-Schwarzschild-Str.~1, 85748 Garching, Germany}
\newcommand{\affilMunich}{Ludwig-Maximilians-Universit\"at M\"unchen (LMU), Geschwister-Scholl-Platz~1, 80539 M\"unchen, Germany}
\newcommand{\affilNanjing}{School of Astronomy and Space Science, Nanjing University, Nanjing 210023, China}
\newcommand{\affilNijmegen}{Department of Astrophysics/IMAPP, Radboud University Nijmegen, P.O.~Box 9010, 6500 GL Nijmegen, The Netherlands}
\newcommand{\affilNikhef}{Nikhef, Science Park Amsterdam, 1098 XG Amsterdam, The Netherlands}
\newcommand{\affilPurpleMt}{Key Laboratory of Dark Matter and Space Astronomy, Purple Mountain Observatory, Chinese Academy of Sciences, No.~10 Yuanhua Road, Nanjing, China}
\newcommand{\affilULB}{Universit\'e Libre de Bruxelles, Science Faculty CP230, B-1050 Brussels, Belgium}
\newcommand{\affilVUB}{Inter-University Institute For High Energies (IIHE), Vrije Universiteit Brussel (VUB), Pleinlaan 2, 1050 Brussels, Belgium}
\newcommand{\affilXidian}{School of Electronic Engineering, Xidian University, No.2 South Taibai Road, Xi'an, China}
\newcommand{\affilSKA}{SKA Observatory, Jodrell Bank, Lower Withington, Macclesfield, SK11 9FT, UK}
\newcommand{\affilIITK}{Department of Physics, Indian Institute of Technology Kanpur, Kanpur, UP-208016, India}

\affiliation[a]{\affilVUB}
\affiliation[b]{\affilNijmegen}

\affiliation[c]{\affilKIT}
\affiliation[d]{\affilErlangen}
\affiliation[e]{\affilManchester}

\affiliation[f]{\affilCurtin}
\affiliation[g]{\affilMaxPlanck}
\affiliation[h]{\affilMunich}
\affiliation[i]{\affilGorlitz}
\affiliation[j]{\affilGroningen}
\affiliation[k]{\affilASTRON}
\affiliation[l]{\affilPurpleMt}
\affiliation[m]{\affilNikhef}
\affiliation[n]{\affilDESY}
\affiliation[o]{\affilKanpur}
\affiliation[p]{\affilKhalifa}
\affiliation[q]{\affilCanTho}
\affiliation[r]{\affilSKA}
\affiliation[s]{\affilNanjing}
\affiliation[t]{\affilKeyNanjing}
\affiliation[u]{\affilXidian}
\affiliation[v]{\affilHefei}
\affiliation[w]{\affilULB}
\affiliation[x]{\affilIITK}

\emailAdd{a.corstanje@astro.ru.nl}

\abstract{Cosmic rays arrive at Earth with energies ranging from $10^9$ to over $\unit[10^{20}]{eV}$. One of the open questions in high-energy cosmic ray science concerns the origin of the highest-energy cosmic rays that can be accelerated by Galactic sources, and the transition energy beyond which only extragalactic sources can provide. Measuring the mass composition gives essential information for comparing measurements to source and propagation models, both from the abundances at the source and from the maximum attainable energy which is proportional to the particle charge (and hence its mass).
The highest-energy cosmic rays from the Galaxy are found in a range of $10^{16}$ to $\unit[10^{18}]{eV}$ which is well suited for radio detection. Building on a decade of experience in measuring cosmic rays at LOFAR, we show that SKA-Low, augmented with an array of small particle detectors, is well suited to advance the field by measuring the mass composition of cosmic rays across this energy range. }

\begin{document}
\maketitle

\section{Introduction}
Cosmic rays arrive on Earth in an energy range spanning about 11 decades, from $10^9$ to over $\unit[10^{20}]{eV}$. Therefore, at the high end of the spectrum we find the most energetic known particles we are able to measure.

\begin{figure}
    \centering
	\includegraphics[width=0.49\columnwidth]{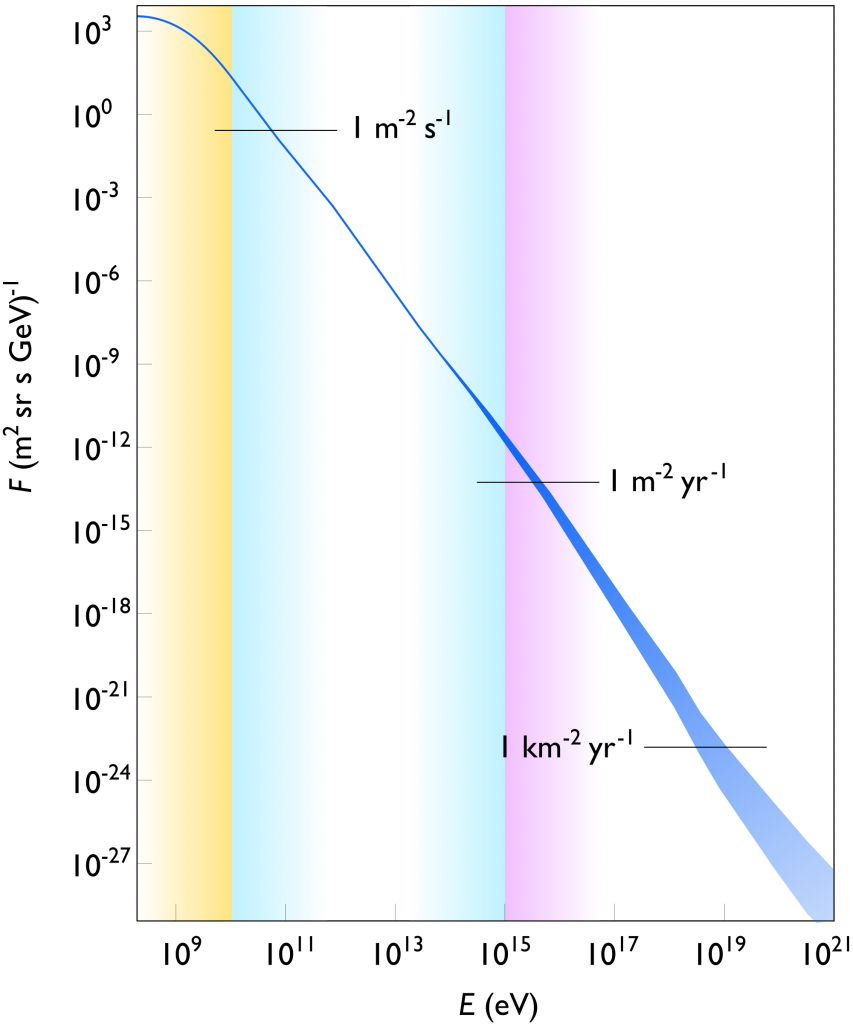}
    \caption{The cosmic-ray energy spectrum. While low-energy cosmic rays are abundant, the flux drops steeply with increasing energy, at a power law of roughly $E^{-3}$ up to smaller features. At the highest energies, very large detector arrays are needed to collect significant data.}
    \label{fig:flux_spectrum}
\end{figure}

The cosmic-ray energy spectrum exhibits a steep fall-off with increasing energy (see Fig.~\ref{fig:flux_spectrum}), making direct detections above about $\unit[10^{14}]{eV}$ infeasible. Instead, cosmic rays at higher energies are detected indirectly through cascades of secondary particles they produce in the atmosphere, called {\it extensive air showers}. 
The particles that reach the ground can be measured in particle detectors.
In addition, the secondary particles collectively emit electromagnetic radiation that is coherent, hence detectable, at radio wavelengths.
A general introduction to cosmic rays including recent results can be found in \cite{ParticleDataGroup:2024cfk} and its update from 2025 in \cite{ParticleDataGroup:2025}.

A simplified estimate of the maximum energy cosmic rays can attain in a given source is given by the Hillas criterion \citep{Hillas:1984},
\begin{equation}
E_{\mathrm{max}} = q B R,
\end{equation}
where $q$ is the charge of the particle, $B$ is the average magnetic field strength, and $R$ is the radius of the source.
From this general consideration, it follows that towards the highest energies, sources of cosmic rays must be very large and/or have high magnetic field strengths.
Moreover, the maximum energy for a given source depends on the particle charge, and therefore on the particle mass as well.
This means that when the cosmic ray spectrum can be split into a mass composition as a function of energy, this encodes information about the sources and their maximum attainable energies.

In the energy range of $10^{16}$ to $\unit[10^{18}]{eV}$, a transition is expected from sources within the Galaxy to extragalactic sources.
This means the high-energy limit attainable by sources within the Galaxy falls in this range, and this limit is lower for light particles such as protons and helium nuclei, than for heavy particles such as iron nuclei.
Thus it is these sources, the strongest accelerators in the Galaxy, that we seek to study using radio measurements as the relatively compact arrays such as LOFAR and SKA-Low have sufficient collecting area to probe this energy region. The Galactic radio background which is the main source of `noise' in the antenna signals, sets the lower energy limit for detection.
The high-energy limit is set by the steeply falling cosmic-ray spectrum combined with the effective area of the detector. These coincide with the transition energy range.

Supernova remnants are candidates for sources able to accelerate to PeV energies (known as PeVatrons), as are young massive star clusters and pulsar wind nebulae (see e.g.~\cite{deOnaWilhelmi:2024mul}). The chapter on PeV gamma rays in this book \citep{SKA_Book_pev_gammas} lists these sources in more detail, with additional references.

\begin{figure}
    \centering
	\includegraphics[width=0.49\columnwidth]{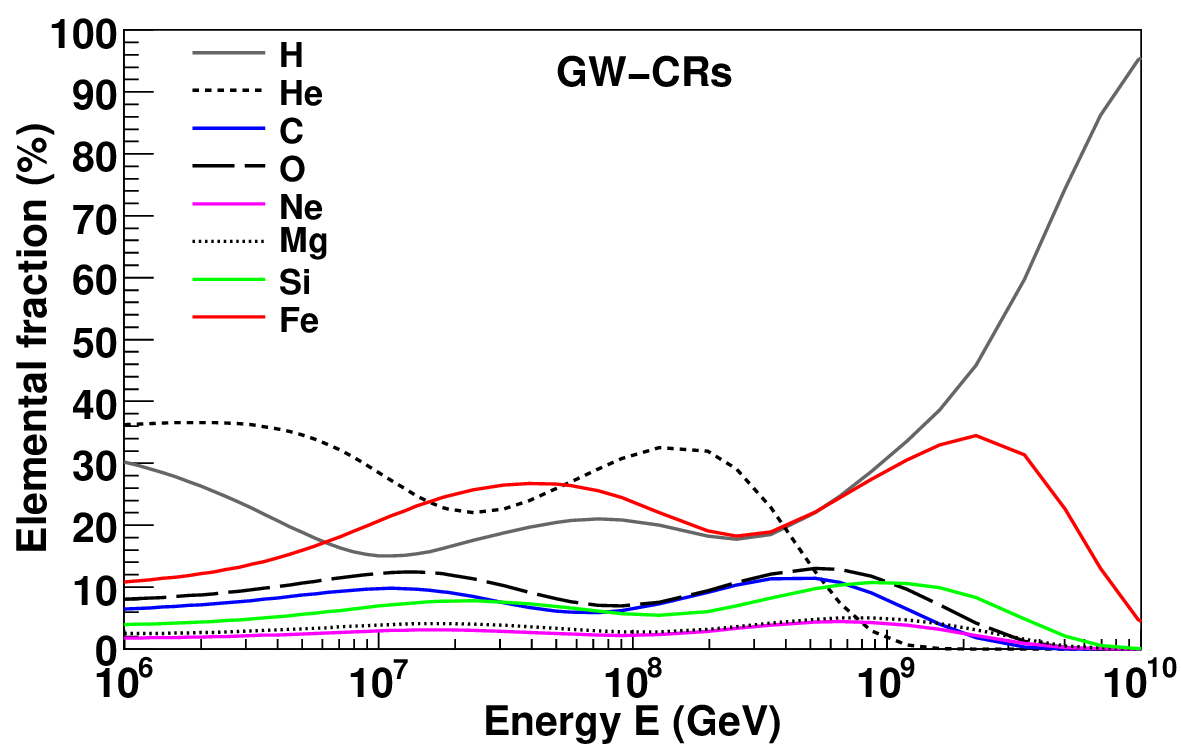}
	\includegraphics[width=0.49\columnwidth]{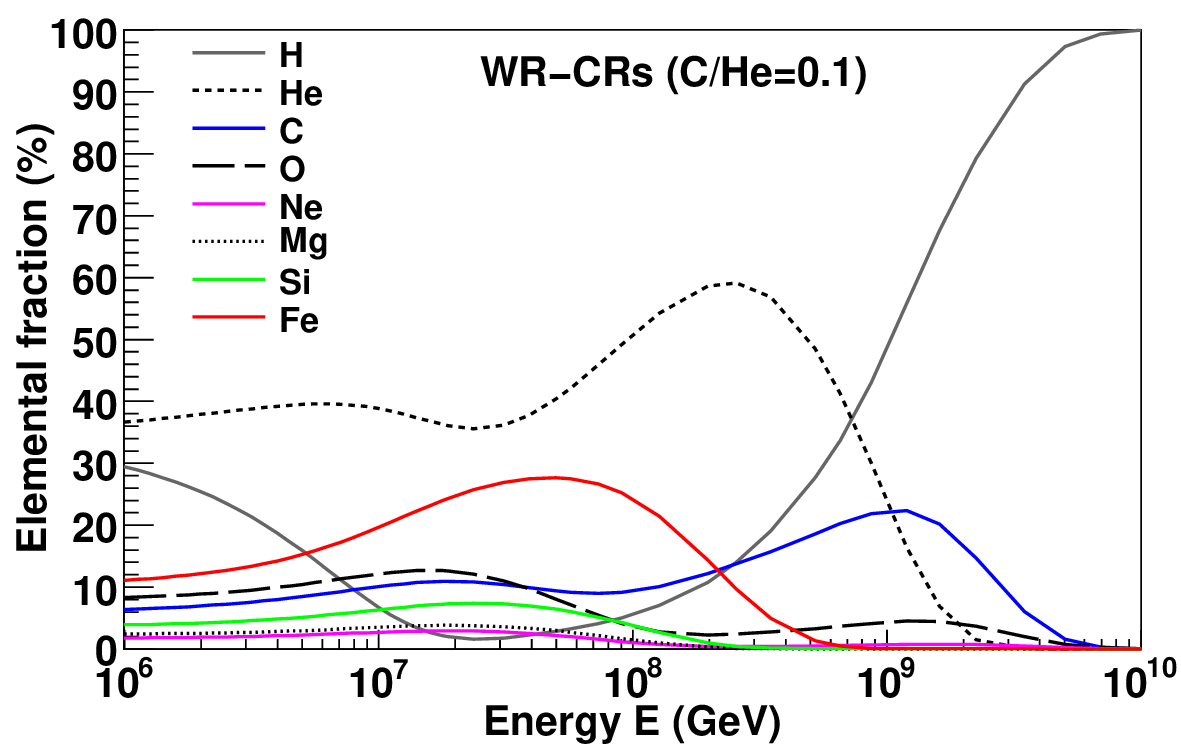}
    \caption{The mass composition fractions for two different scenarios of Galactic cosmic ray production at the highest energies attainable. They are labeled GW for Galactic wind, and WR for Wolf-Rayet supernovae. A notable difference in hydrogen/helium ratios is visible. The flux cuts off around $\unit[10^9]{GeV} = \unit[10^{18}]{eV}$, hence the abundance fractions are only reliable up to this energy. Beyond this, mainly extragalactic protons remain. }
    \label{fig:fractions_wind_WR}
\end{figure}

A complete picture of high-energy cosmic ray production is still elusive, and multiple scenarios are considered.
For instance, in \cite{Thoudam:2016} it is argued that the most energetic particles from the Galaxy may arise from re-acceleration of particles meeting the Galactic termination shock, or shockwave acceleration in supernovas expanding into strongly magnetized winds around Wolf-Rayet stars. 
The mass composition of cosmic rays as arising from these sources has been modeled, producing the curves in Fig.~\ref{fig:fractions_wind_WR}. They show considerable differences in mass composition, especially in the ratio of hydrogen to helium. It was further shown that these sources may explain the cosmic-ray flux up to energies of about $\unit[10^{18}]{eV}$, before cutting off and extragalactic cosmic rays dominate. 

Dedicated cosmic-ray observatories have been built around the world, each having their specific strengths.
The largest experiment, Pierre Auger Observatory in Argentina \citep{Abraham:2004}, featuring water Cherenkov detectors and a sparse array of radio antennas, is able to see the highest-energy cosmic rays, up to $\unit[10^{20}]{eV}$ and slightly above, as its area of about $\unit[3000]{km^2}$ allows to measure these rare particles in a reasonable observing time.
The Telescope Array \citep{TALE:2020} in Utah, USA also has a large collecting area of $\unit[762]{km^2}$, and its low-energy extension is able to measure air showers down to below $\unit[10^{16}]{eV}$ of primary energy.

\begin{figure}
    \centering
	\includegraphics[trim={0cm 0cm 11.0cm 0cm},clip,width=0.49\columnwidth]{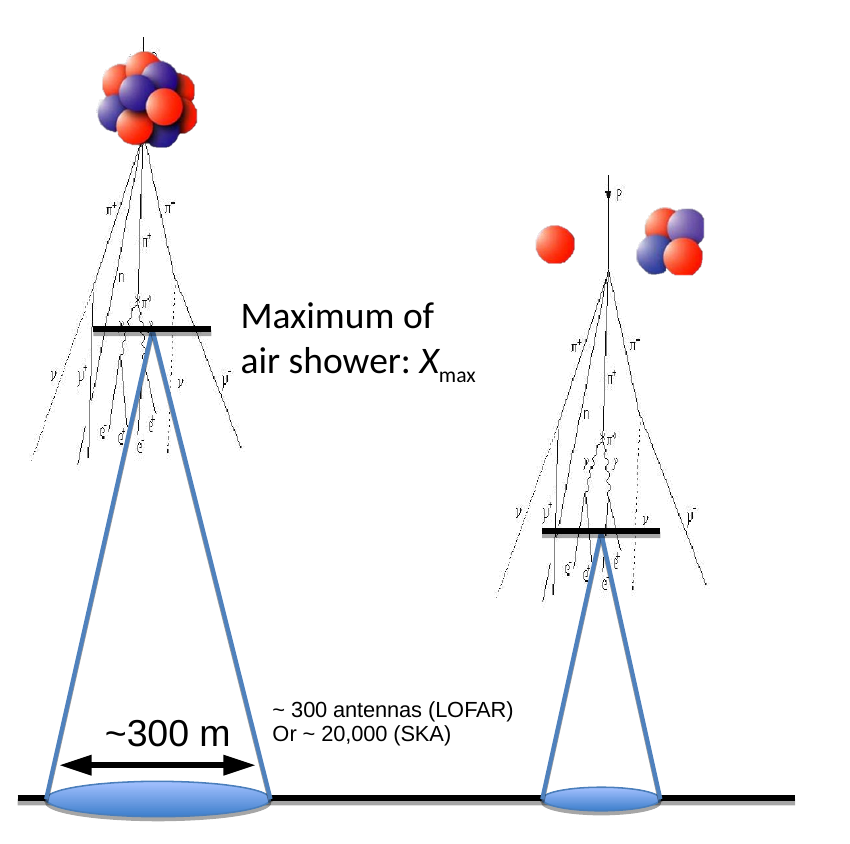}
        \includegraphics[width=0.49\columnwidth]{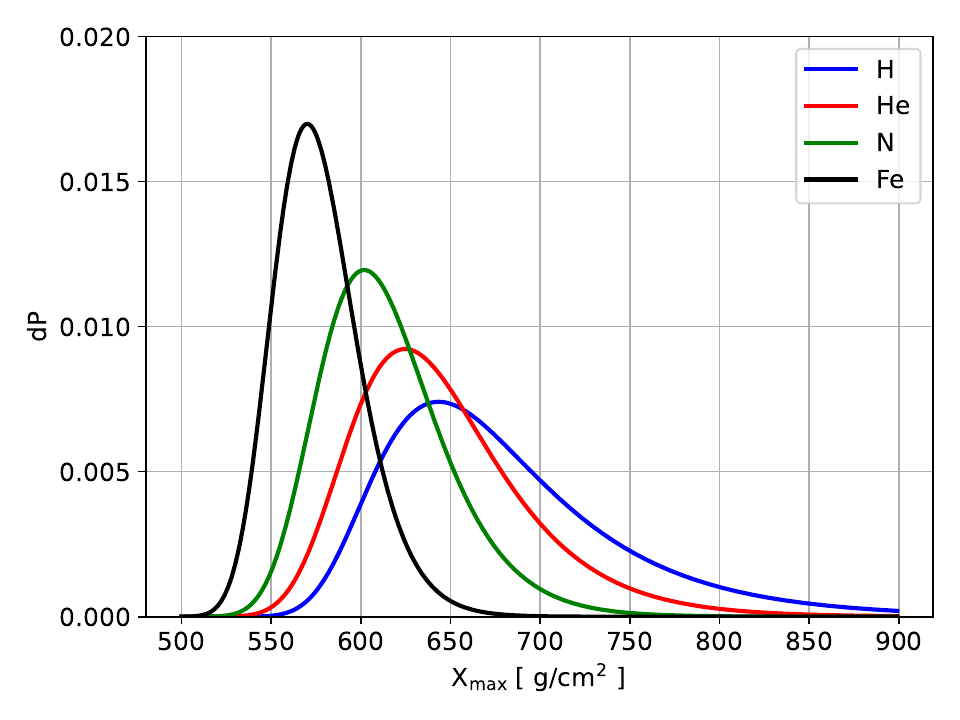}
    \caption{Left: a schematic picture showing how, to lowest order, the radio footprint becomes smaller for air showers that reach a maximum \Xmax closer to the ground (higher values), allowing to infer \Xmax from data. Right: probability densities of \Xmax for four selected primary elements; nitrogen is usually chosen as proxy for C/N/O.}
    \label{fig:xmax_distributions}
\end{figure}

The radio detection method can be used as a stand-alone method using a relatively small particle detector array as trigger, such as done at LOFAR over the past decade \citep{Schellart:2014oaa,Thoudam:2014cfa,Buitink:2016nkf}. This project has demonstrated the benefits of an observing mode in the background of a distributed radio telescope.
The collecting area below $\unit[1]{km^2}$ limits the measurements to a lower energy than the large experiments, but its much larger antenna density allows for detailed measurements of lower-energy air showers, and important observables like the shower maximum and the primary energy are measured to an accuracy in line with the state of the art. The shower maximum is defined as the point in the atmosphere where the number of particles is maximal, and denoted \Xmax. To lowest order, the radio footprint size depends on \Xmax as depicted in Fig.~\ref{fig:xmax_distributions} (see further discussion in Sect.~\ref{sect:measuring}).

In general, fluorescence detection is the most direct way to track the progression of the air shower, imaging its faint fluorescence trail using a telescope \citep{Kampert:2012}. However, it requires dark nights with clear skies, which strongly limits its duty cycle and needs to operate far away from light pollution.
The radio method has none of these constraints, but infers the air shower evolution more indirectly.

Another use of the radio method is to augment existing particle detector and fluorescence detection, such as done at Pierre Auger Observatory with the AugerPrime array upgrade \citep{AugerPrime:2019} which features a radio antenna alongside each particle detector, at a spacing of $\unit[1.5]{km}$ throughout the array, with more dense subsections as well.

Measuring cosmic rays with the SKA-Low core region would follow the same principles as LOFAR, at an even much higher antenna density as well as a larger collecting area. This opens up new capabilities aimed at high-precision measurements of individual air showers, surpassing any current methods. Combining antennas using interferometry or similar techniques opens up a new, lower energy range previously unexplored by radio detection. This is described below in Sect.~\ref{sect:beamforming} and explored further in Sect.~\ref{sect:interferometry}.

\subsection{Cosmic-ray measurements at LOFAR}
In the context of the present work, the LOFAR radio telescope can be seen as a precursor for cosmic-ray measurements at SKA-Low.
The Low Frequency Array (LOFAR) \citep{vanHaarlem:2013}, like SKA-Low, is a distributed radio telescope consisting of omnidirectional antennas. The core region in the north of the Netherlands features nearly 600 low-band antennas (LBA) in an area of $\unit[320]{m}$ diameter, half of which could be used simultaneously in the past. The frequency range of 30 to $\unit[80]{MHz}$ is suitable for detecting cosmic ray signals.

In the antenna fields, a small particle detector array has been placed \citep{Thoudam:2014cfa}, comprising 20 scintillator detectors which are flat boxes of about $\unit[1]{m^2}$ area. They serve mainly as a trigger for reading out transient buffers containing raw digitized voltages from each antenna dipole.
Triggering on radio pulses only is possible in principle, but there are considerable practical challenges, such as the need for dedicated hardware that is able to reject millions of pulse trains that may happen at times in an environment with radio-frequency interference.

To have a particle detector array (or any electronic devices) located at the site of an operational radio telescope, it is essential that the detectors are designed and tested to produce negligible radio-frequency interference (RFI). To this end they were tested in a radio anechoic chamber located at ASTRON and found to comply with the requirements for LOFAR.

Since detecting the first cosmic ray in 2011, LOFAR has measured a variety of properties of the radio `footprint' of air showers in high detail, among which the curved wavefront in which the signals arrive \citep{Corstanje:2014waa,Apel:2014usa}, its polarization signature \citep{Schellart:2014oaa}, and the pulse energy footprint that is the basis for current mass composition measurements \citep{Buitink:2014eqa,Buitink:2016nkf}. The most recent mass composition analysis \citep{Corstanje:2021kik} confirmed the significant low-mass component in the energy region of interest.

\section{Measuring cosmic rays with a radio telescope}\label{sect:measuring}
In the following sections we show the essential steps of the analysis of a cosmic-ray measurement as it would be done at SKA-Low. This includes example time series of the signal pulses, as they would be read out from a transient buffer recording raw ADC voltages per antenna dipole. Measuring the energy fluence in the pulses of all buffered antennas in the SKA-Low core with significant signal, a {\it radio energy footprint} arises. This is the basis of a model-data comparison, as radio footprints can be constructed from simulated air showers plus the SKALA4 antenna model as implemented in the NuRadioMC software \citep{Glaser:2019cws,NuRadioMC_code}. The parameters of interest in the best-fitting showers, which are known in simulations, give a reliable estimate for the measured shower.

Measuring the cosmic-ray mass composition is mainly based on the {\it depth of shower maximum}, called \Xmax, expressed as atmospheric column depth in $\unit{g/cm^2}$. On average, low-mass particles penetrate deeper into the atmosphere before interacting, producing shower maxima closer to the ground (higher \Xmax values). This is depicted schematically in Fig.~\ref{fig:xmax_distributions}, along with the probability distributions for \Xmax for selected primary particles at energy $\unit[10^{17}]{eV}$.
It is seen that in this rather simplified picture, the radio footprint size depends on the distance to the shower maximum by geometry.
Secondary parameters describing the shower development also carry information on the primary particle, as shown in Sect.~\ref{sect:higherorder}.

We also discuss how offline beamforming of the data increases signal-to-noise ratios (SNR) at low primary energies, which helps to considerably extend the measurable energy range downwards.

Results on \Xmax shown below are based on the detailed simulation study in \cite{Corstanje:2025} which makes use of the analysis techniques developed for LOFAR \citep{Buitink:2014eqa,Buitink:2016nkf,Corstanje:2021kik}.

\begin{figure}
    \centering
	\includegraphics[width=\columnwidth]{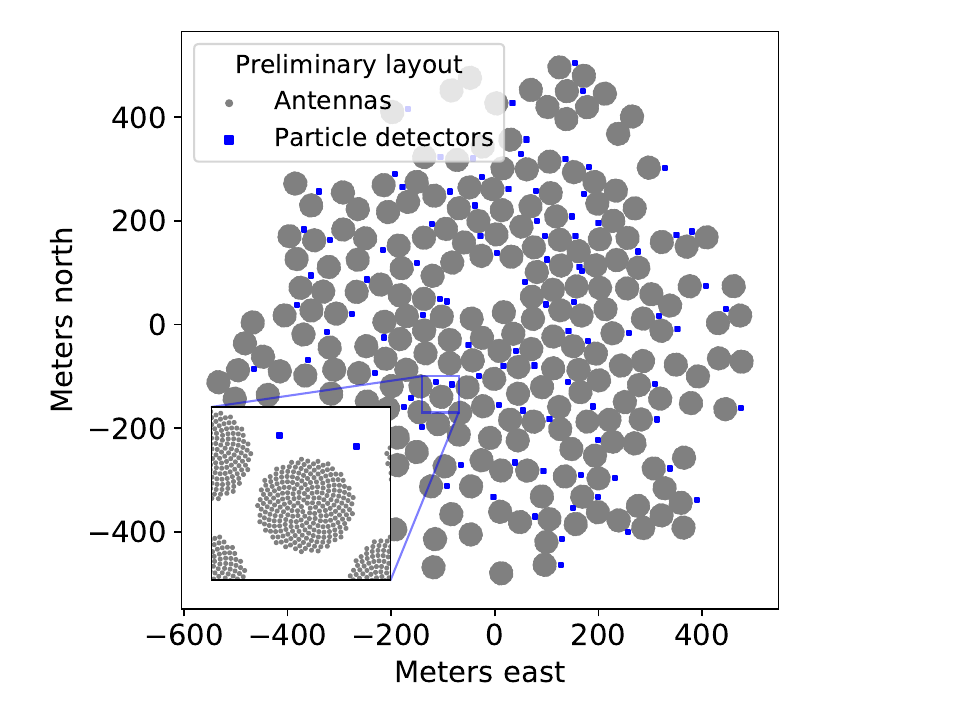}
    \caption{The antenna layout of the SKA inner core at the AA4 stage, with an example distribution of 100 particle detector boxes of $\unit[1]{m^2}$ (points not to scale). The final locations are to be optimized to on-site logistics. The earlier development milestone denoted AA* comprises about $\unit[90]{\%}$ of this array. Figure taken from \cite{Corstanje:2025}.}
    \label{fig:ska_layout}
\end{figure}

\subsection{The raw material: buffered voltage time traces}
The radio pulses from cosmic ray air showers are quite short, on the order of 10 to $\unit[100]{ns}$. Capturing and measuring signals well in the sub-microsecond range requires having access to the raw digitized voltage data per antenna. After typical pre-processing done in a distributed radio telescope, such as polyphase filtering and/or beamforming, too much time resolution is lost.
Therefore, a cosmic-ray observing mode is based on reading out raw ADC voltages from a transient buffer per antenna at the time a cosmic ray signal arrives. 

Triggering a buffer readout is best done using an on-site particle detector array, as this gives proof that a cosmic ray was the source of the measured signal, plus near-real-time information on the air shower that can be used for the trigger logic. Moreover, a radio-only trigger would be challenged by impulsive RFI. This would require considerable real-time processing on dedicated hardware to keep the number of false positives under control, while having to deal with occasional bursts of many short RFI pulses.

Using a scintillator-based particle detector array alongside an operational radio telescope is an established technique and has been demonstrated at LOFAR over the past decade \citep{Thoudam:2014cfa,Thoudam:2015lba}.
An example layout of about 100 particle detectors, consisting of flat boxes about $\unit[1]{m^2}$ in area, is depicted in Fig.~\ref{fig:ska_layout}, along with the AA4 layout of the inner core region. As the AA* subset of this array already contains about $\unit[90]{\%}$ of the full array, this will already be sufficient for carrying out the complete cosmic-ray science operations. The exact locations of the particle detectors can be varied to suit the on-site possibilities such as cabling positions. The distribution should be roughly uniform where possible, but there is flexibility to align with on-site logistical constraints.

Triggering a buffer readout requires the trigger signal from the particle detector array to freeze the buffers within a time limit set by the buffer length. This sets an important constraint to the total response time of the system, probably dominated by network turn-around time.
After this, the buffer is read out and the data is processed offline. Reading out the buffer is less time-sensitive, although it counts as dead time for transient observing modes based on the raw antenna signal buffers.

\begin{figure}
    \centering
	\includegraphics[width=0.7\columnwidth]{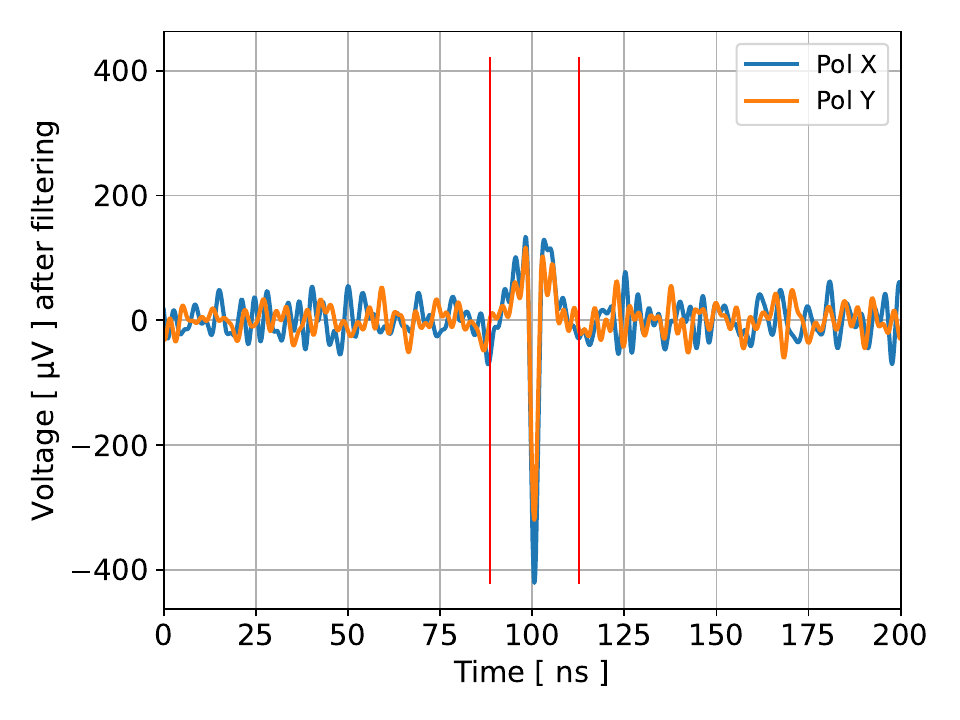}
    \caption{A voltage trace at one of the antennas in Fig.~\ref{fig:ska_layout} where the signal is strong. Using the NuRadio software \citep{Glaser:2019cws}, the SKALA4 antenna model is applied to the simulated electric field traces, producing the voltage at the antennas. Background noise from the Galaxy has been added, and the vertical lines indicate a time window in which the pulse energy is estimated. A filter has been applied to compensate for dispersion from the antenna characteristic, producing a sharp, short pulse. Figure taken from \cite{Corstanje:2025}.}
    \label{fig:example_trace}
\end{figure}

An example time trace of one antenna is shown in Fig.~\ref{fig:example_trace}, which shows a pulse (after a de-dispersion filter) from a simulated $\unit[10^{17}]{eV}$ proton air shower, plus Galactic noise. 
The vertical bars indicate a time window in which to measure the pulse energy.

\begin{figure}
    \centering
	\includegraphics[width=0.8\columnwidth]{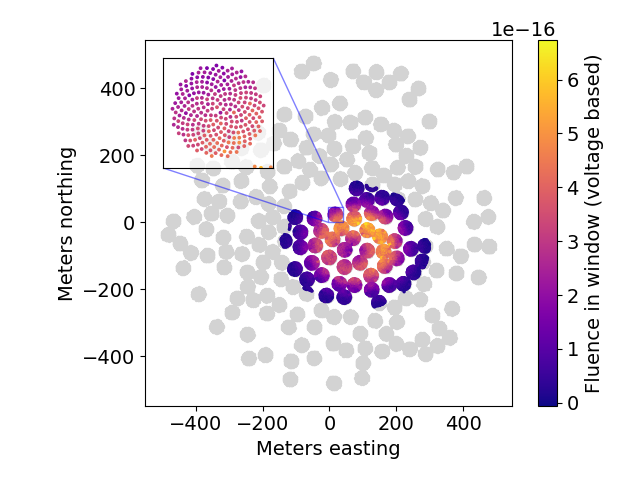}
    \caption{An example of a cosmic-ray radio footprint from a $\unit[10^{17}]{eV}$ proton. The color-code shows the energy fluence in the pulses at each antenna. A detection threshold was set at 5 sigma. Figure taken from \cite{Corstanje:2025}.}
    \label{fig:footprint_example}
\end{figure}

After measuring the pulse energy in both dipoles of all buffered antennas, we obtain the radio energy footprint, an example of which is shown in Fig.~\ref{fig:footprint_example}. This lateral distribution of radio energy is compared to an ensemble of air shower simulations, spanning a range of primary particle masses and \Xmax values. 

\subsection{Reconstructing the shower maximum from the measured radio footprint}\label{sect:reco}

Having established the radio footprint, measured by the antennas in the SKA-Low inner core, we can compare this to radio footprints from simulated air showers.
To this end, we use CORSIKA \citep{Corsika:1998} and CoREAS \citep{CoREAS:2013} which simulate individual particles and their interactions as they pass through the atmosphere, to produce their combined radio signal as electric fields at the antenna locations. 
Applying the antenna model to the electric fields (and in a later stage also other parts of the signal chain as filters), we obtain the voltage traces, from which the pulse energies are measured using the same procedure as for data.

To measure the depth of shower maximum \Xmax, we use an ensemble of simulated showers that spans the natural \Xmax range, and fit the simulated energy fluences to the measured ones. This produces a chi-squared fit quality which is optimized to produce the best fit:
\begin{equation}
\chi^2 = \sum_{\mathrm{antennas}} \left(\frac{A\,f_{\mathrm{model}}(x-x_0, y-y_0) - f_{\mathrm{data}}(x, y)}{\sigma_f(x, y)}\right)^2,
\end{equation}
where an overall scale factor $A$ and the core position $(x_0, y_0)$ are free parameters in the fit, and $f_{\mathrm{model}}$ and $f_{\mathrm{data}}$ are the energy fluences per antenna from simulations and data, respectively. The uncertainty on the energy fluence measurements at each antenna is given by $\sigma_f(x, y)$.

\begin{figure}
    \centering
	\includegraphics[width=0.49\columnwidth]{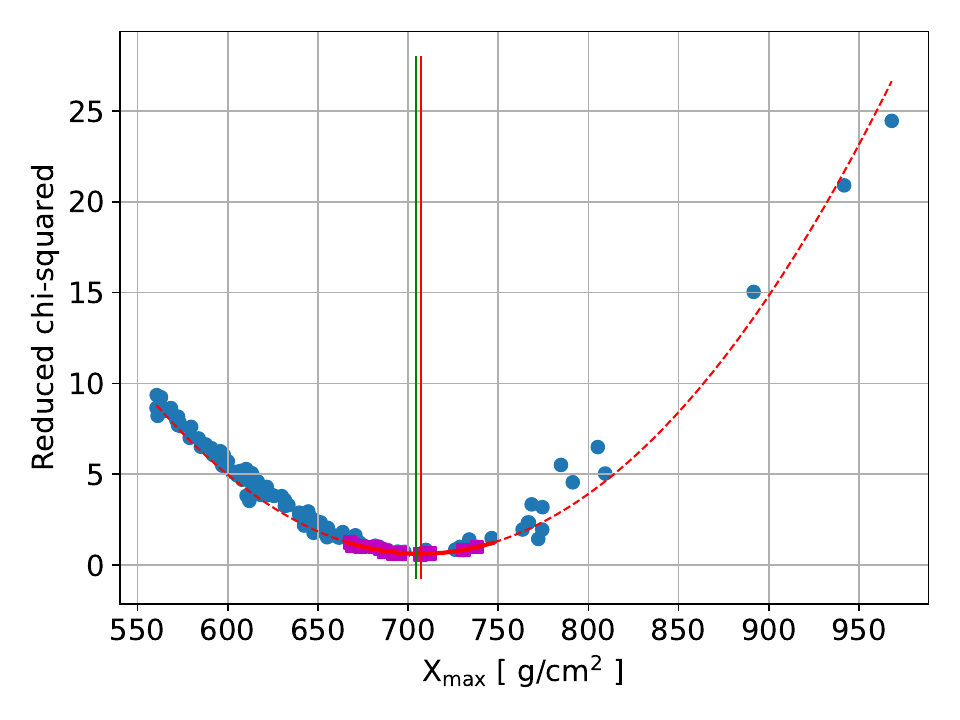}
        \includegraphics[width=0.49\columnwidth]{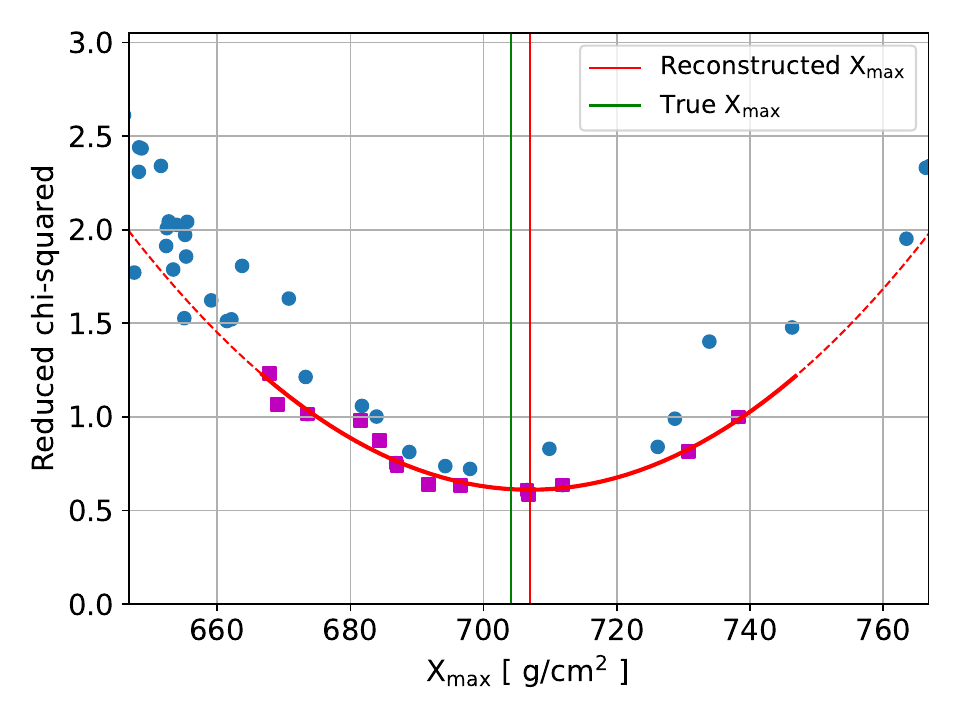}
    
    \caption{An example of an \Xmax reconstruction. Fitting each of the showers in the ensemble gives a reduced $\chi^2$ value, which follows a parabolic curve near its minimum value. The minimum of a parabola fitted to the lower envelope of points is taken as the \Xmax estimate. The right panel shows a close-up. Figure taken from \cite{Corstanje:2025}.}
    \label{fig:reco_example}
\end{figure}

The best-fitting shower gives an estimate of \Xmax in the measured shower.
To overcome limitations from a finite density of simulated showers, we fit a parabola through the lower envelope of $\chi^2$ points as a function of \Xmax, in a range close to the minimum, as shown in Fig.~\ref{fig:reco_example}.
The minimum of the parabola is taken as the \Xmax estimate.

\subsection{Using offline beamforming of groups of nearby antennas to boost signal-to-noise ratios}\label{sect:beamforming}
With so many antennas in a dense array, it is natural to use beamforming, for instance to boost the signal-to-noise ratio.
In a cosmic-ray air shower, most of the radio emission arises in a region in the atmosphere centered on the shower maximum \Xmax.
Typical distances of \Xmax to the antenna array are on the order of 3 to $\unit[7]{km}$ for air showers that arrive at zenith angles up to roughly $\unit[45]{deg}$; very inclined showers arriving from close to the horizon are not discussed here. As a result, the pulse arrives at the array as a curved wavefront \citep{Corstanje:2014waa,Apel:2014usa}. Therefore, approaches based on far-field beamforming as generally used in astronomy are not optimal. 
Approaches based on near-field beamforming to infer \Xmax exist but obtaining unbiased estimates is not straightforward \citep{Scholten:2024,Schoorlemmer:2020low}. Further preliminary results based on this approach are described in Sect.~\ref{sect:interferometry}.

We circumvent these issues for now by making use of the high antenna density of SKA-Low, noting that on length scales of roughly $\unit[10]{m}$, the wavefront curvature is negligible. Hence, combining groups of e.g.~$N_{\mathrm{ant}}=4$, 16, or 64 antennas in one station by far-field beamforming perpendicular to the wavefront, we can boost the SNR by a factor $\sqrt{N_{\mathrm{ant}}}$. 
In the simulation study we have emulated this process by reducing the number of antennas by a factor $N_{\mathrm{ant}}$, and increasing the SNR by the corresponding square-root factor.

\subsection{Accuracy in reconstructing the shower maximum}
We have done the \Xmax reconstruction as outlined in Sect.~\ref{sect:reco}, taking in turn each of the 140 showers in the ensemble for each geometry as mock data. This way, comparing reconstructed to true \Xmax values, we can gather statistics on the reconstruction accuracy. 
We have assumed that overall, 1 in 4 antennas of the array will have buffered data available.

\begin{figure}
    \centering
	\includegraphics[width=0.49\columnwidth]{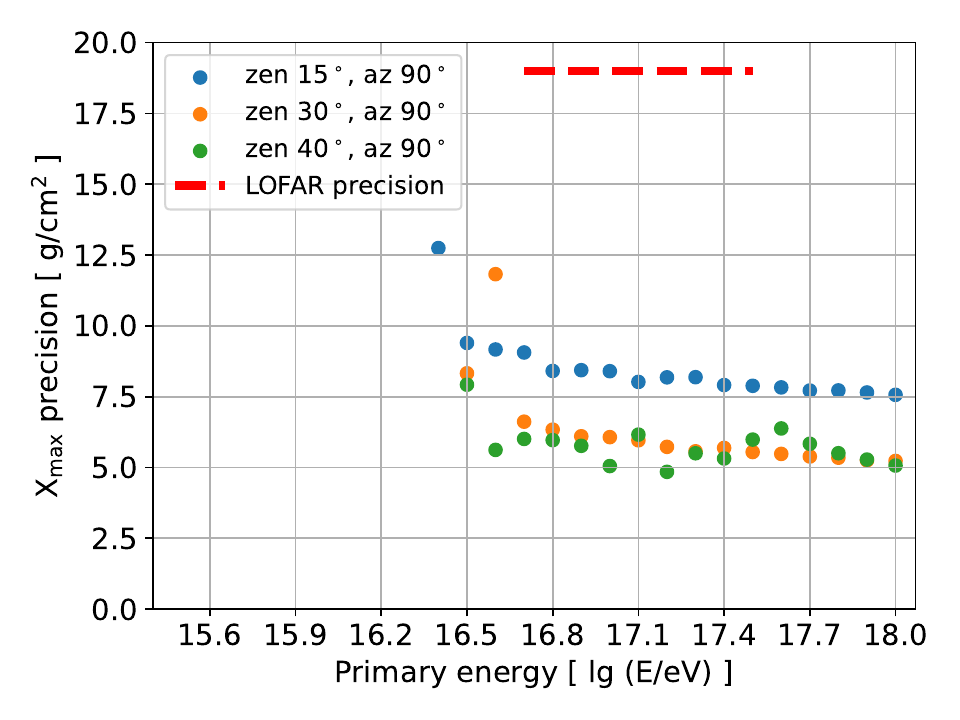}
        \includegraphics[width=0.49\columnwidth]{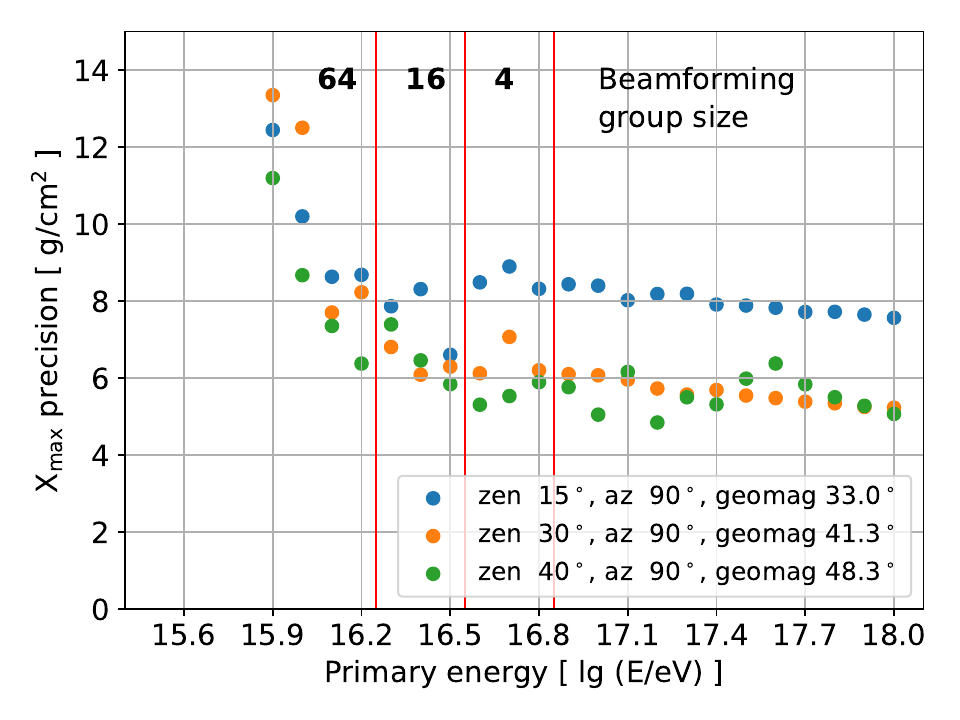}
    
    \caption{The precision in reconstructing \Xmax, versus primary energy. Left panel: using single antennas. Right panel: using beamforming in groups of 4, 16, or 64 antennas, with higher numbers towards lower primary energy. This considerably lowers the reliable detection threshold. Figure taken from \cite{Corstanje:2025}.}
    \label{fig:xmax_precision}
\end{figure}

The results for the precision are summarized in the left panel of Fig.~\ref{fig:xmax_precision}, where signals in each antenna have been used separately. An \Xmax precision of 15 to $\unit[20]{g/cm^2}$ is considered state of the art in the field, achieved by fluorescence detection. Radio detection at LOFAR performs at this level as well.
Notably, SKA-Low will perform nearly a factor 3 better for individual air showers. In fact, we find that the remaining errors are no longer limited by the number of antennas or by the signal-to-noise ratios. Instead, they reflect variations in the air shower evolution independent of \Xmax, that have been neglected here. We further discuss this below in Sect.~\ref{sect:longdist}. We have also evaluated the bias due to the reconstruction steps, which is generally below $\unit[1.5]{g/cm^2}$. This is quite acceptable given a total systematic uncertainty budget around $\unit[10]{g/cm^2}$. However, as measurements with SKA-Low naturally raise the bar for accuracy, any source of bias needs close attention. Future improvements of the reconstruction methods which are being developed will address this.

The low-energy cutoff for single-antenna detection is around $\unit[10^{16.5}]{eV}$; below this level, too few antennas reach the detection threshold (here set to five sigma above the noise) for a reliable reconstruction.

We have evaluated the expected improvements using the group-wise beamforming method outlined in Sect.~\ref{sect:beamforming}, see the right panel of Fig.~\ref{fig:xmax_precision}. We find that using progressively larger antenna groups towards lower energy, the reliable detection threshold is lowered to about $\unit[10^{15.9}]{eV}$. This opens up a considerable energy range previously unavailable to radio detection. Moreover, as cosmic rays are much more abundant at lower energies, a good mass composition estimate as a function of primary energy is made possible within a reasonably short observation time.

\section{Expected event rates for mass composition estimates}
As seen in Fig.~\ref{fig:xmax_distributions}, the probability distributions of the shower maximum for different primary elements overlap. Moreover, the high-energy hadronic interaction model that is used causes small but significant variations in these distributions.
Hence, to have a good mass composition analysis, a tight systematic uncertainty budget is required, as well as a sufficient number of measured showers.
The mass composition estimate amounts to inferring which linear combination of the given curves (i.e.~the mix fractions) fits the data best. The curves also change slightly with \Xmax uncertainty, and with energy, mainly shifting to higher \Xmax for higher energies; this can be accounted for on a per-shower basis in a maximum likelihood analysis.

The number of measurements is limited by nature at high energies, as cosmic rays become rare and the effective area of SKA-Low's inner core requires long observing times.
On the other hand, at the low end of our energy range, cosmic rays are abundant, and data collection will be limited by technical and operational constraints. 

The systematic uncertainties will always be finite, and as a consequence there is a dataset size (order of magnitude) for which the statistical errors become negligible compared to the systematic errors.
Thus, collecting more measurements in that energy bin would not lead to a better mass composition estimate -- at least when using this method based on \Xmax only. Progressing to more advanced methods, more data may again be advantageous.

\begin{figure}
    \centering
	\includegraphics[width=0.49\columnwidth]{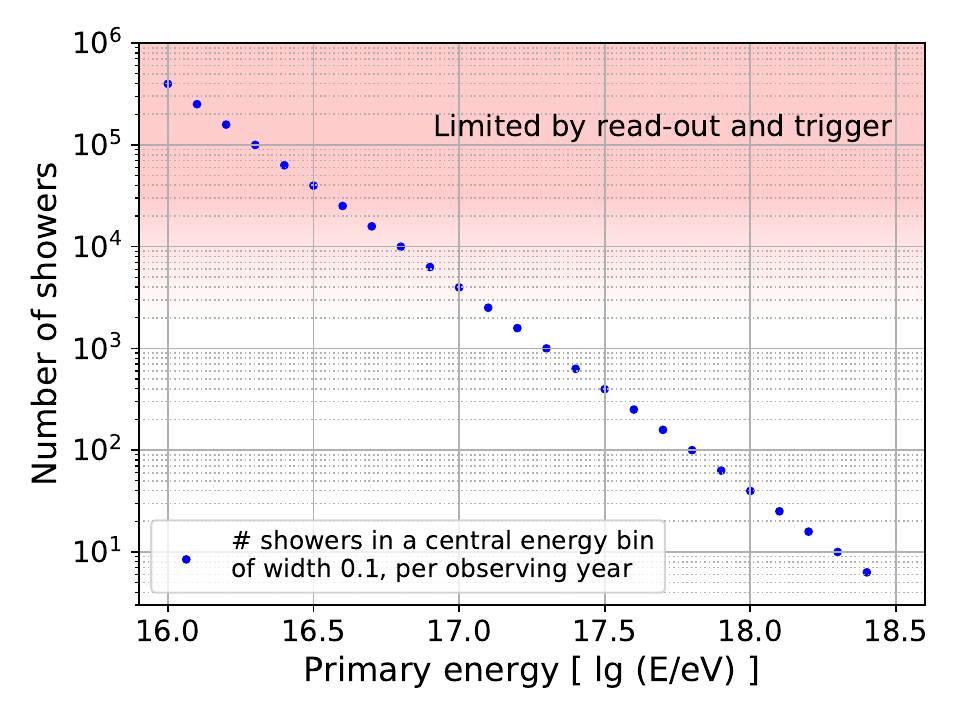}  
    \includegraphics[width=0.49\columnwidth]{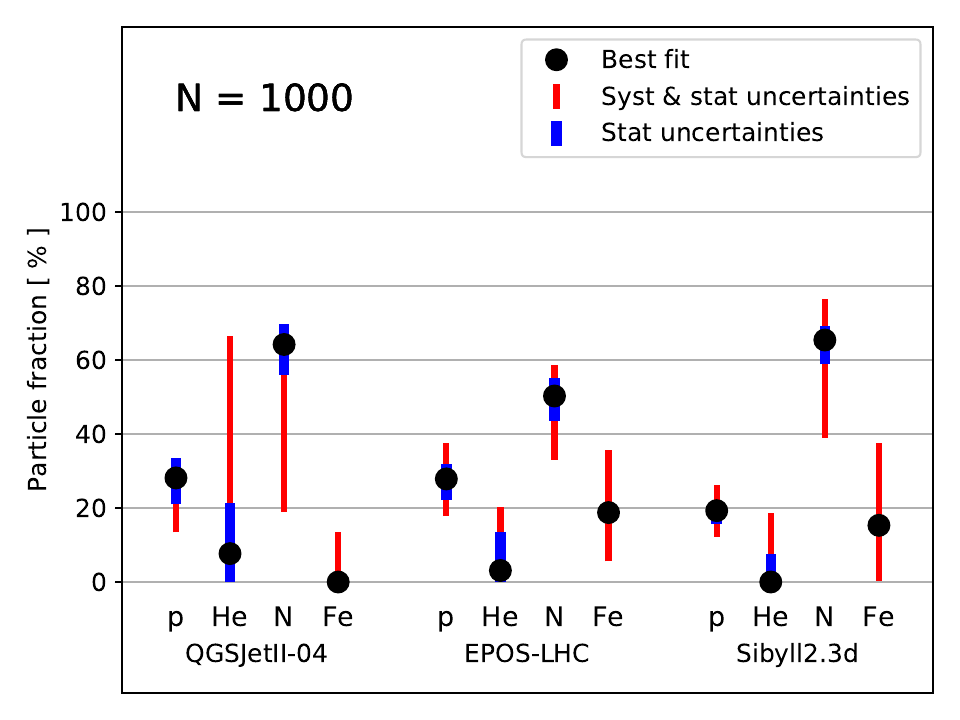}   
    \caption{Left: The expected number of measured air showers in bins of width 0.1 in log-energy, in one net observing year. Technical limitations are not yet known, but are expected around an order of magnitude indicated by the area shaded in red. Right: example uncertainties on the mass composition fractions in a mock dataset of 1000 showers, assuming the same systematic uncertainties as for the LOFAR analysis. Figure taken from \cite{Corstanje:2025}.}
    \label{fig:event_counts}
\end{figure}

In Fig.~\ref{fig:event_counts} (left panel) we have plotted an approximate cosmic ray spectrum, expressed in number of showers per energy bin, per net observing year with SKA-Low. We have chosen a narrow energy bin width of 0.1 in log-energy as used by large cosmic ray observatories, as seeing detailed trends with energy is desired.

To put these numbers into perspective, in the right panel we show example uncertainties as they are found from a bootstrapped \Xmax dataset of size $N=1000$, using the best-fit composition found at LOFAR as a starting point. This would be for a single energy bin as from the left panel of this figure. This happens to be the optimal order of magnitude for a dataset in one energy bin; systematic uncertainties were copied from the LOFAR analysis \citep{Corstanje:2021kik} as they are not expected to change much. The results for three widely used hadronic interaction models reflect the variations in \Xmax distributions they produce.
We see that below an energy of about $\unit[10^{17.3}]{eV}$, 1000 measured showers per energy bin are reached within one year of net uptime of the cosmic-ray observing mode.
At higher energies, wider energy bins or longer observing times will be needed, while reasonable results can already be obtained at lower statistics levels.

In order to limit the burden on the buffering system, the network bandwidth etc., it will likely be necessary to rate-limit the measurements of low-energy showers. A quick, preliminary energy estimate from the particle detectors may serve as input for the trigger decision logic.
As an example, limiting the number of showers per energy bin to 1000 per year of uptime would amount to a trigger rate of roughly twice per hour on average. This seems quite reasonable, as a similar agreement of triggering once per hour was made at LOFAR.

The (uncompressed) data size for one measured shower would scale as
\begin{equation}
    D_{\mathrm{shower}} = \unit[1.2]{GB}\, \left(\frac{f_{\mathrm{sampling}}}{\unit[800]{MHz}}\right)\,\left(\frac{\Delta t_{\mathrm{trace}}}{\unit[50]{\mu s}}\right)\,\left(\frac{N_{\mathrm{ant}}}{15000}\right),
\end{equation}
for sampling rate $f_{\mathrm{sampling}}$, signal trace length $\Delta t$, and number of antennas $N_{\mathrm{ant}}$ centered on about $15000$ for 1 in 4 antennas of the inner core region. In particular the signal trace length is tunable to suit our needs as well as the network and  local storage constraints.
Thus, a cumulative data volume per observing year would scale as
\begin{equation}
    D = \unit[21]{TB}\, \left(\frac{t_{\mathrm{obs}}}{\unit[1]{yr}}\right)\,\left(\frac{f_{\mathrm{trigger}}}{\unit[2]{hr^{-1}}}\right)\,\left(\frac{\Delta t_{\mathrm{trace}}}{\unit[50]{\mu s}}\right)\,\left(\frac{N_{\mathrm{ant}}}{15000}\right),
\end{equation}
for a trigger rate currently estimated at twice per hour. This rate amounts to 17500 events per year, which is satisfactory to have at least 1000 showers in each energy bin as in Fig.~\ref{fig:event_counts} where the cosmic ray spectrum allows this within one net observing year.

We conclude that for cosmic-ray science the requirements in terms of data throughput per hour as well as dataset sizes are relatively modest compared to typical astronomical datasets envisioned for SKA-Low.

\section{Towards measuring the full air shower evolution}\label{sect:higherorder}

As shown above, the shower maximum \Xmax is at present the most important observable used to estimate the mass composition of cosmic rays.
But the full longitudinal shower evolution, of which \Xmax is the maximum, contains more information on the primary particle.
We have seen from Fig.~\ref{fig:xmax_precision} that the precision of the \Xmax estimates reaches a plateau at higher energies, i.e.~it does not improve anymore with SNR or number of antennas. Instead, the uncertainty reflects information in the footprint that has been neglected when focusing on \Xmax only.
An instrument like SKA-Low, aiming to reach the highest precision, would be naturally suited to reach further improvements by measuring additional parameters beside \Xmax.

Here we show results of an early investigation into these higher-order parameters, demonstrating that (a) SKA-Low can measure these in individual air showers, using present methods based on pulse energy fluence only, and (b) that new, independent information on mass composition becomes available, in particular for disentangling the hydrogen and helium fractions which are astrophysically the most relevant.

\subsection{The longitudinal distribution and measuring its parameters}\label{sect:longdist}
\begin{figure}
    \centering
	\includegraphics[width=0.49\columnwidth]{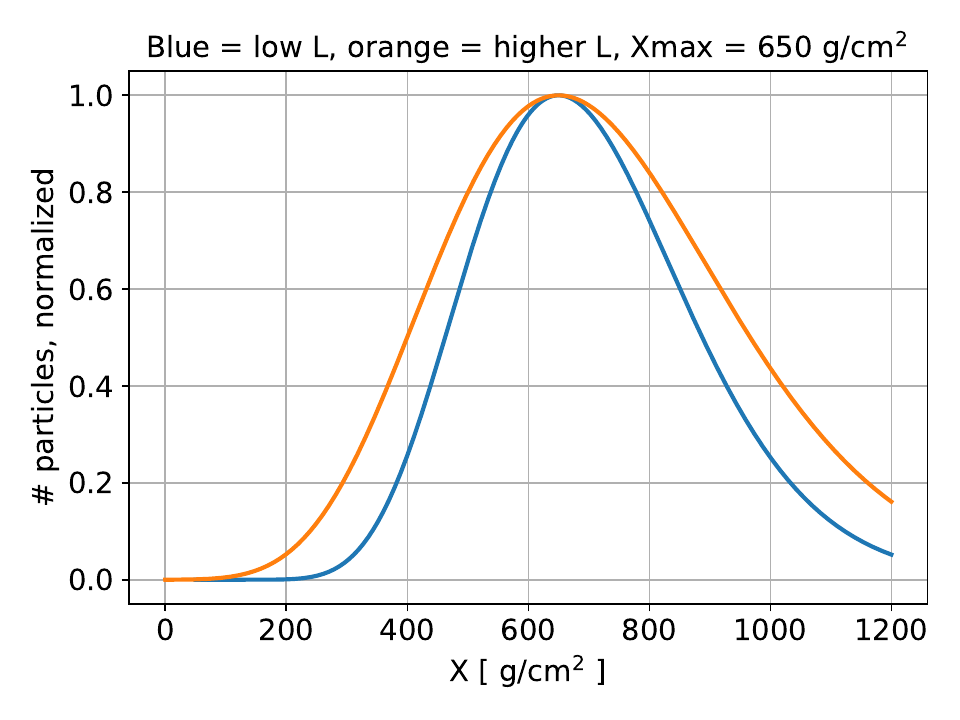}  
    \includegraphics[width=0.49\columnwidth]{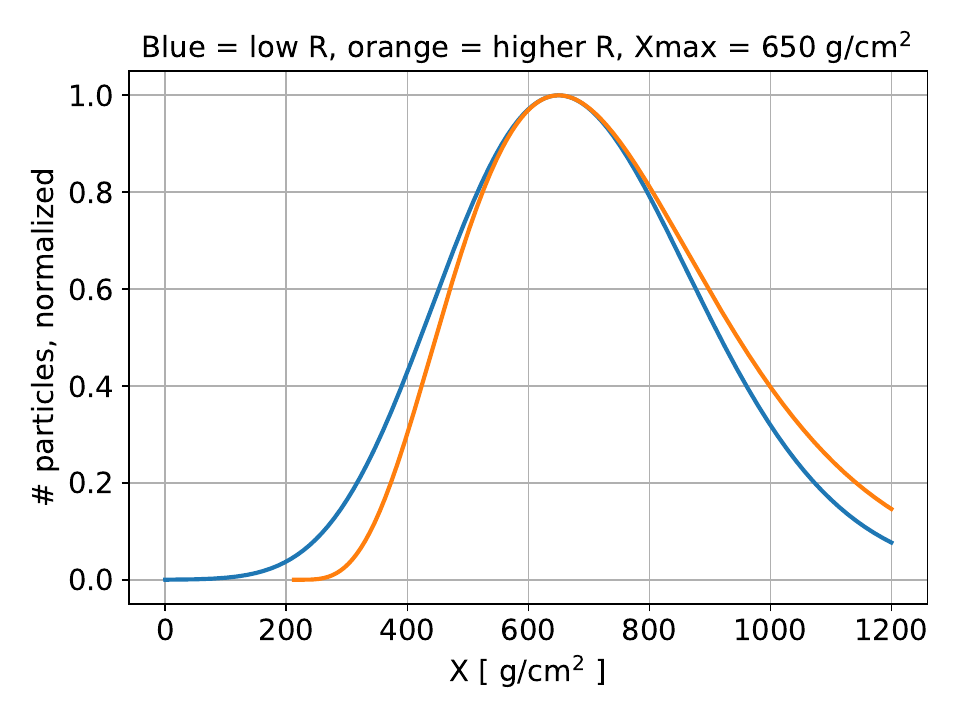}   
    \caption{An example of a typical longitudinal distribution curve. Left: the effect of varying parameter $L$. Right: varying parameter $R$.}
    
    \label{fig:long_distribution}
\end{figure}

Up to some outlier showers, the number of particles as a function of atmospheric column depth can be parametrized \citep{Andringa:2011} by a three-parameter curve as shown in Fig.~\ref{fig:long_distribution}, with a functional form
\begin{equation}\label{eq:long_parametrization}
N(X) = \exp\left(-\frac{X - \Xmaxmath}{RL}\right)\,\left(1 + \frac{R}{L}\left(X - \Xmaxmath\right)\right)^\frac{1}{R^2}.
\end{equation}
It specifies the relative number of particles $N(X)$, normalized to unity at $X=\Xmaxmath$, and the parameters $L$ and $R$ are proportional to the variance and skewness of the distribution, respectively.

There exist showers that do not conform to this curve. They mainly arise when in the first few interactions, an energetic particle is produced that travels relatively long before interacting again, producing effectively a second shower further down in the atmosphere.
Comprising roughly 0.5 to $\unit[1]{\%}$ of air showers, they can feature a double-peak curve in the evolution of the number of particles. As these showers carry information about the high-energy hadronic interactions, depend on the primary particle, and are measurable and distinguishable by SKA-Low, they are of interest in their own right. They are discussed further in \cite{SKA_Book_hadr_physics}.

We have tested whether radio fluence footprints are sensitive to these higher-order parameters $L$ and $R$, and if they are measurable using the same technique as used for \Xmax (for more details see \cite{Corstanje:2023uyg}).
To this end we have created an ensemble of 110 showers all having the same shower maximum $\Xmaxmath=645 \pm \unit[0.5]{g/cm^2}$, by pre-selecting showers with the fast CONEX simulations \citep{Bergmann:2007} that quickly produce only a longitudinal evolution. 

From a direct comparison of (noiseless) radio footprints, we found that the primary sensitivity is on a combination of the $L$ and $R$ parameter. A simple way of combining these is a linear function, given the same dimension as $L$ ($R$ is dimensionless), as in
\begin{equation}\label{eq:LRcombi}
S(L, R) = L + \frac{\unit[16]{g/cm^2}}{0.06} \left(R - 0.3\right),
\end{equation}
where the coefficients that give maximal sensitivity depend on the frequency bandwidth, and probably on the incoming direction and \Xmax as well. In this case, we chose a 50 to $\unit[100]{MHz}$ bandwidth filter to have a relatively strong weight on $L$; for higher frequencies, the sensitivity appears to lean more towards $R$.
This dependence, and a possibly more suitable (re)parametrization are still under investigation.

\begin{figure}
    \centering
	\includegraphics[trim=0cm 0 14.12cm 0, clip, width=0.7\columnwidth]{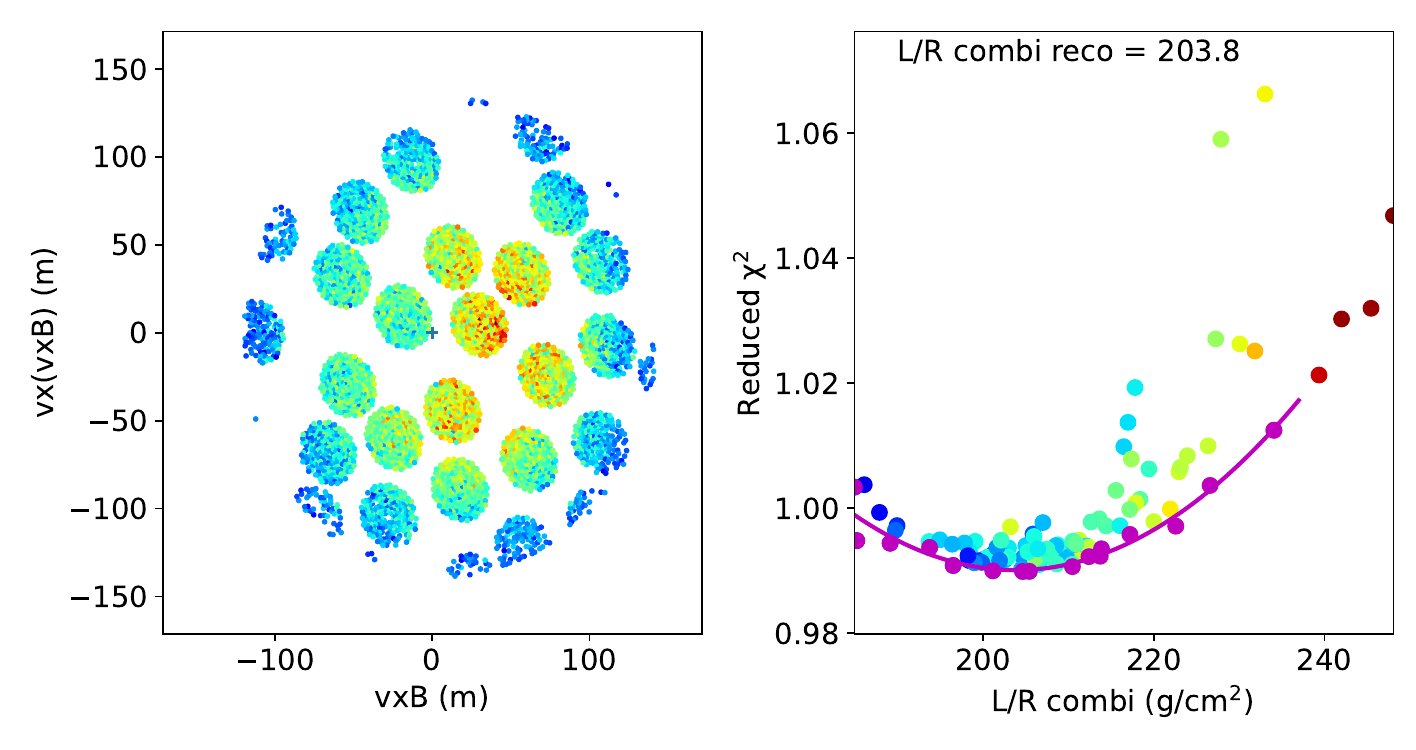}  
    \caption{Measuring the sensitive parameter $S$ combining curve parameters $L$ and $R$ from Eq.\ref{eq:LRcombi} using the same fitting procedure as for \Xmax. A clearly detectable optimum near the true value is found, whereas the small variations in reduced chi-squared values indicate that a considerable number of antennas is needed to measure this accurately. Color-coding of the data points in the right panel is proportional to $L$, and the magenta points indicate the lower envelope used for fitting the parabola. Figure taken from \cite{Corstanje:2023uyg}.}
    \label{fig:measure_S}
\end{figure}

To demonstrate the measurement of $S$, we have followed the same procedure as for \Xmax described above, albeit for an older version of the antenna model and noise model. The results are shown in Fig.~\ref{fig:measure_S}.
The optimum is found near the true value of $\unit[204.6]{g/cm^2}$, and the fitted parabola closely follows the lower envelope of points near the minimum.

Thus, this preliminary investigation has shown that SKA-Low will be sensitive to higher-order parameters beyond \Xmax, already using the present methods based on pulse energy fluence only.
This is only a lower bound to what can be achieved, as for instance methods that employ the full pulse traces, such as interferometry or information field theory \citep{SKA_Book_ift}, are expected to leverage the extra information in the full pulse signals.

In the next subsection, we explore how in particular the parameter $L$ gives additional information on mass composition. The results carry over to a combination of $L$ and $R$, as long as there is enough sensitivity on $L$.

As the parameter space becomes multi-dimensional, the requirements for the number of simulations to match to the data scale up accordingly.
This may become prohibitive when the number of measured showers is very large, such as anticipated in the SKA era. Thus, strategies are called for to drastically speed up this process. 

One such approach, recently developed and described in \cite{Desmet:2025}, uses a simulated shower as a base to create radio footprints for showers with a variety of longitudinal profiles. 
This speeds up the process by orders of magnitude, allowing to narrow down all longitudinal parameters of a measured shower. Its application to such parameter inference is still under development.

Limiting the parameter space before applying `expensive' simulations is another approach, for instance by doing an initial fit for \Xmax using a footprint parametrization \citep{Nelles:2015} or a fast simulation method \citep{Scholten:2018} and only then use the more time-consuming and accurate CoREAS simulations to achieve the final accuracy and obtain the higher-order parameters. This requires pre-selecting simulated showers to have the desired parameter values, which using current versions of Corsika is possible with CONEX.

Another alternative approach is based on the geometrical reconstruction of the radio emission profile of air showers by backtracking the antenna  signals at the ground to their emission points along the shower axis \citep{Vuta:2025}. The method is computationally highly efficient as it requires minimal input from simulations, and has the potential to reconstruct the shower \Xmax directly from the observed data.

\subsection{Mass composition information from shower evolution parameters}
To see the dependence of \Xmax and $L$ in particular on the primary particle mass, we have simulated an ensemble of CONEX showers, at primary energies $10^{16}$, $10^{17}$, and $\unit[10^{18}]{eV}$ respectively \citep{Buitink:2023reh}. These are fast simulations (a few minutes per shower) that produce only the longitudinal evolution, to which we fit the three-parameter curve from Eq.~\ref{eq:long_parametrization}.
Each ensemble has 2500 showers for each of 5 primary particle elements, and the simulations were repeated for three hadronic interaction models, EPOS-LHC \citep{EPOSLHC:2013}, QGSJetII-04 \citep{Ostapchenko:2013}, and Sibyll-2.3d \citep{Riehn:2019jet}.

\begin{figure}
    \centering
	\includegraphics[width=0.7\columnwidth]{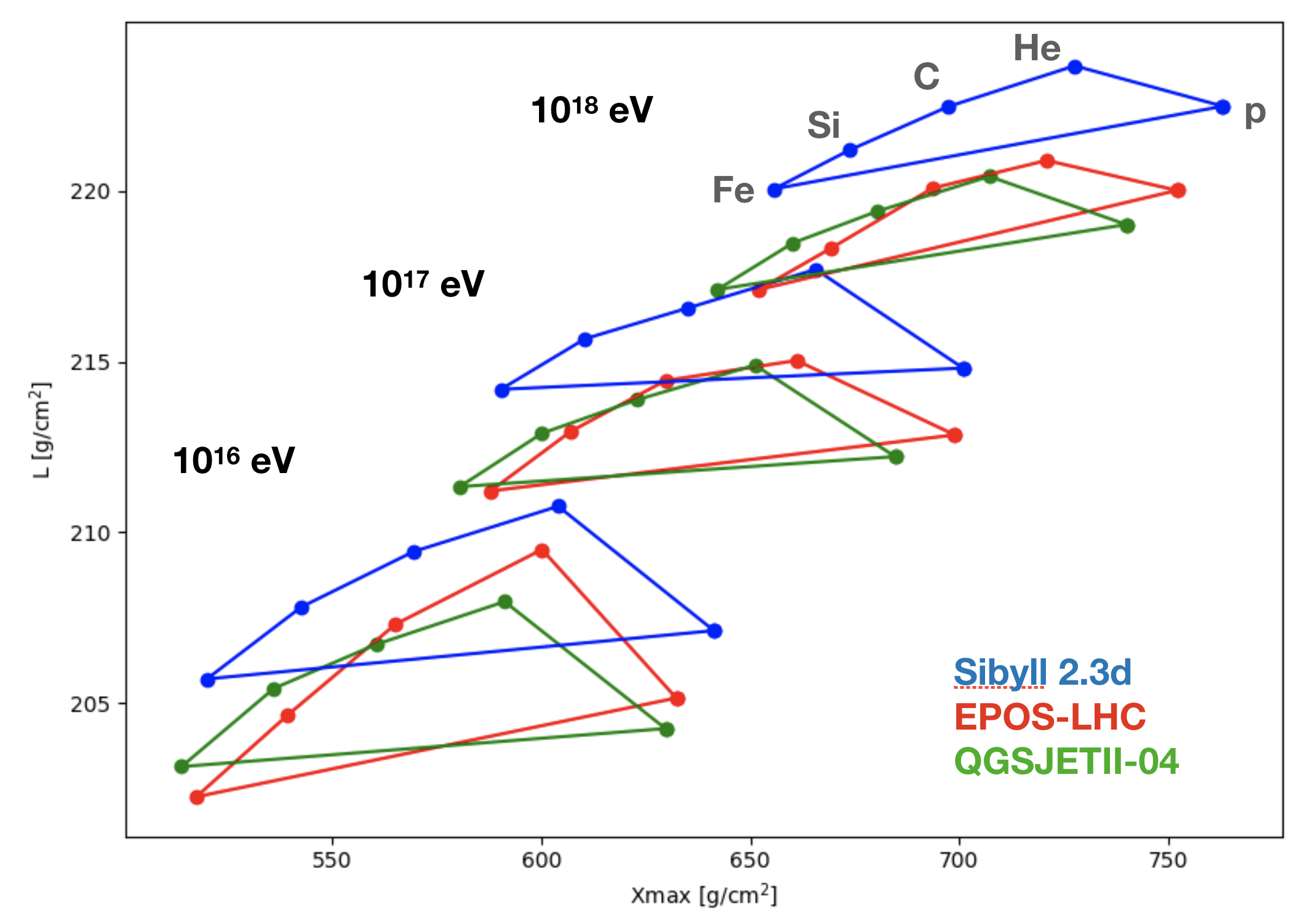}  
    \caption{The mean values of $L$ and \Xmax for four different elements, at primary energies $10^{16}$, $10^{17}$, and $\unit[10^{18}]{eV}$, and for three hadronic interaction models. Notably, hydrogen (noted as p, for protons) stands out from the other elements, which is helpful for mass composition analysis. Figure taken from \cite{Buitink:2023reh}.}
    \label{fig:triangle_plot_L}
\end{figure}

Plotting the mean value of $L$ for each element versus the mean \Xmax, we obtain Fig.~\ref{fig:triangle_plot_L}. We see that hydrogen (protons) is different from the other elements as it has a notably lower mean $L$ than for instance helium, the next light element.
A dataset of showers with a mixed mass composition will have average values inside the indicated triangles.

\begin{figure}
    \centering
	\includegraphics[width=0.7\columnwidth]{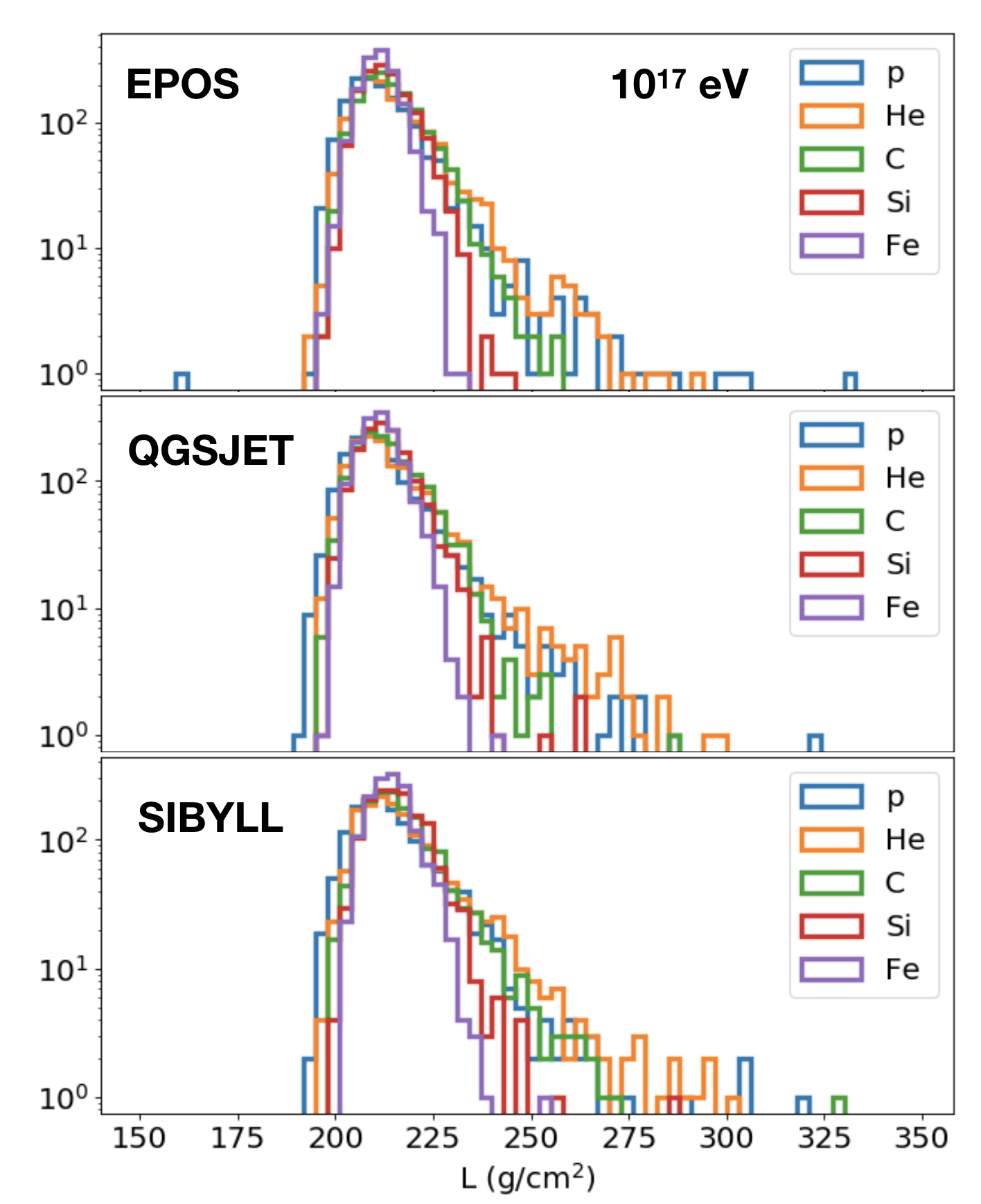}  
    \caption{Histograms of $L$ for various elements, with raw counts on the vertical axis; the tail end becomes longer towards lighter elements, up to helium. Again, hydrogen stands out with somewhat shorter tails than hydrogen, which is another piece of information that helps distinguishing hydrogen from helium in a mass composition analysis. Figure taken from \cite{Buitink:2023reh}.}
    \label{fig:distribution_tail_L}
\end{figure}

The mean values depend on energy and on the hadronic interaction model, which may put limits on the accuracy of using the average $L$ for inferring the mass composition. For instance, there are systematic uncertainties on energy of about $\unit[15]{\%}$.
However, there is more information in the measurements, considering the full distribution of a parameter such as $L$. It features a relatively long tail towards higher values, which is much more robust against systematic uncertainties than the average.
The tail end is longest for helium, getting shorter with heavier primary elements.
Again, hydrogen is somewhat outlying as it has a somewhat shorter tail than helium.
The histograms of $L$ are depicted in Fig.~\ref{fig:distribution_tail_L}.

In \cite{Buitink:2023reh} it was shown that a basic approach counting the fraction of tail-end showers with $L > \unit[225]{g/cm^2}$ can be used to infer the hydrogen fraction to roughly $\pm \unit[10]{\%}$, without using the \Xmax distributions or other information. This is a first, important proof of concept.
It is clear that this procedure can be optimized further, for instance by employing a maximum likelihood analysis per individual shower as was done for \Xmax \citep{Corstanje:2021kik}, as well as considering the joint distribution of \Xmax and $L$. Thus, adding measurements of $L$ provides additional, independent information on the mass composition. In particular this helps to distinguish hydrogen from helium, which is important for connecting to astrophysical source models, thus overcoming the difficulty at any dataset size when using only \Xmax, caused by systematic uncertainties. 

\section{On reconstructing the shower maximum using offline near-field interferometry}\label{sect:interferometry}

In Sect.~\ref{sect:reco}, we described a method to reconstruct \Xmax using the pulse energy distribution in the footprint. This method produces robust \Xmax reconstructions, however, it only uses a fraction of the information contained in the radio pulses. 
To make fuller use of the information available, analysis methods utilizing information field theory for near-field interferometry are currently being developed \citep{SKA_Book_ift}. With the same aim, we are currently exploring a near-field beamforming, or interferometric technique \citep{Schoorlemmer:2020low, Schlüter_2021}. In Sect.~\ref{sect:beamforming}, we described a group-wise beamforming method, in which the signals from nearby antennas are combined, boosting the signal-to-noise ratio at lower energies. This provides a way to extend the reconstruction technique described in Sect.~\ref{sect:reco} to lower energies. In this section, we discuss the prospects for near-field, interferometric reconstruction of air shower development at SKA-Low.

In contrast to classical radio interferometry in the far-field, the source of radio emission from a cosmic ray induced extensive air shower is located relatively near the observation site ($\sim$10 km in case of near-vertical incidence), it is not a point-like emitting source, and it does not have a plane wavefront, not even a spherically symmetric wavefront \citep{Corstanje:2014waa, Apel:2014usa}. Antennas that are illuminated in the radio footprint see emission from the entirety of the air shower. On the Cherenkov cone, emission from all parts of the shower arrive at the same time, boosting the signal but complicating the source localisation. A promising technique for interferometric air shower reconstruction, known as the Radio Interferometric Technique (RIT) \citep{Schoorlemmer:2020low,Schluter:2021egm} beamforms the signals from antennas to given locations in the atmosphere, $\vec{j}$, as 

\begin{equation}
    B_j(t) = \sum_i^\text{ant} S_i (t-\Delta_{i,j}).
    \label{eq:beam}
\end{equation}

\begin{figure}
    \centering
	\includegraphics[width=1\columnwidth]{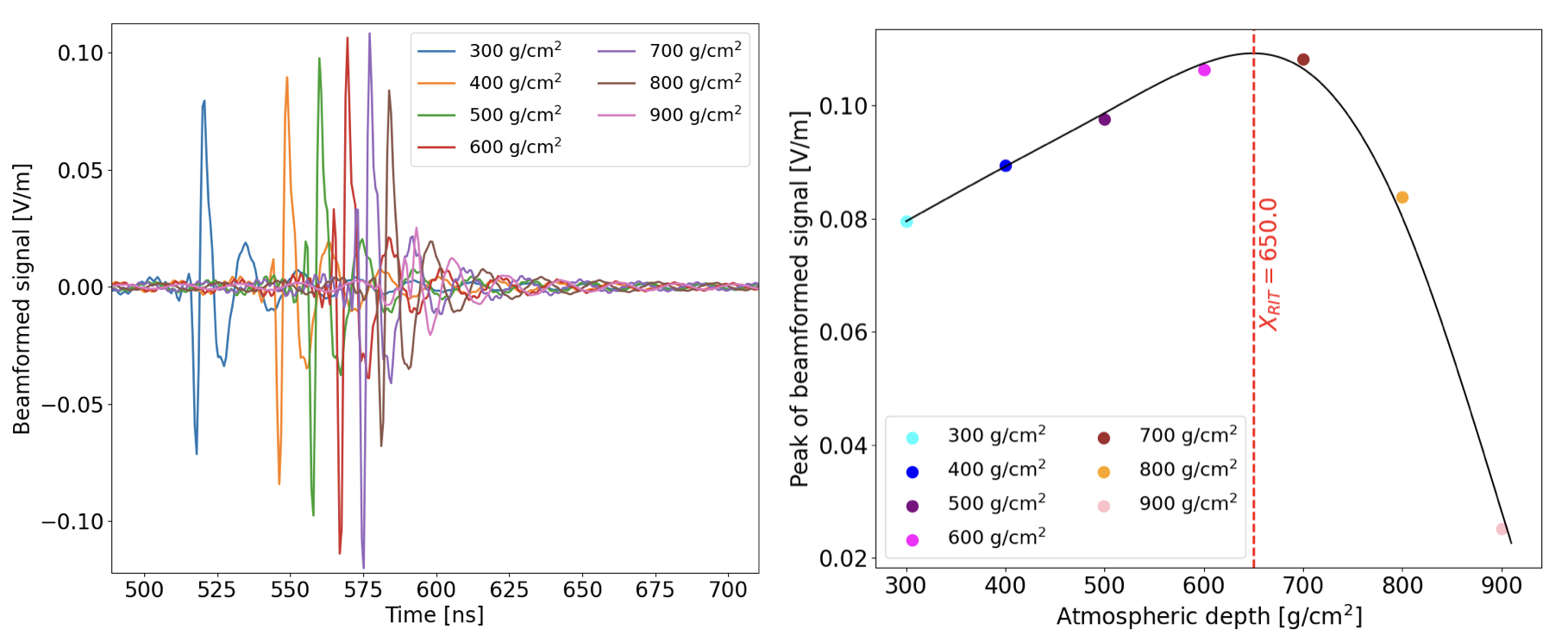}  
    \caption{An example of an interferometric reconstruction of a simulated SKA event. Left: The beamformed signal at varying atmospheric depths along the shower axis. Right: The peak fo the beamformed signal at different atmospheric depths along the shower axis.}
    \label{fig:RIT_example}
\end{figure}

Here $S_i$ are the signals in individual antennas, $\Delta_{i,j}$ is the time-shift according to the antenna location, and $B_j(t)$ gives the strength of the beamformed signal. When signals are beamformed to a point close to the main shower emission region, the signals add coherently and result in a larger combined signal. The RIT method reconstructs a peak signal at a geometric point referred to as \Xrit. \Xrit is strongly correlated with the traditional \Xmax, with details of the correlation depending on the antenna layout, frequency range of interest, and air shower geometry. The RIT method is very promising not only for reconstructing \Xmax, but also finer details of air shower development as well as enhancing signals from low energy air showers. A demonstration of the  interferometric technique as applied to the SKA-Low AA* antenna configuration is shown in Figure~\ref{fig:RIT_example}. The left panel shows the beamformed signal at varying atmospheric depths along the shower axis, with different colored lines indicating different depths. The right panel shows the corresponding power in the beamformed signal with a clear peak at \Xrit.

SKA-Low has great promise for performing interferometric air shower reconstructions. The RIT method requires excellent (sub nanosecond level) time synchronization between antennas, which will be the case for SKA-Low. Additionally, the technique works best with sufficient antenna coverage across the entire radio footprint, which will be the case at SKA-Low. Finally, the sheer number of antennas that can be included in the interferometric reconstruction will allow us to enhance very small signals, thereby reconstructing low energy air showers (see the discussion of PeV gamma-ray event reconstruction in \citealp{SKA_Book_pev_gammas}). The potential to have a consistent reconstruction method for air showers over the $10^{15} - 10^{18}$~eV range will be valuable for producing cosmic-ray composition measurements across the entire transition. This is the necessary input to distinguish different source classes of the highest energy Galactic cosmic rays.

\begin{figure}
    \centering
	\includegraphics[width=1\columnwidth]{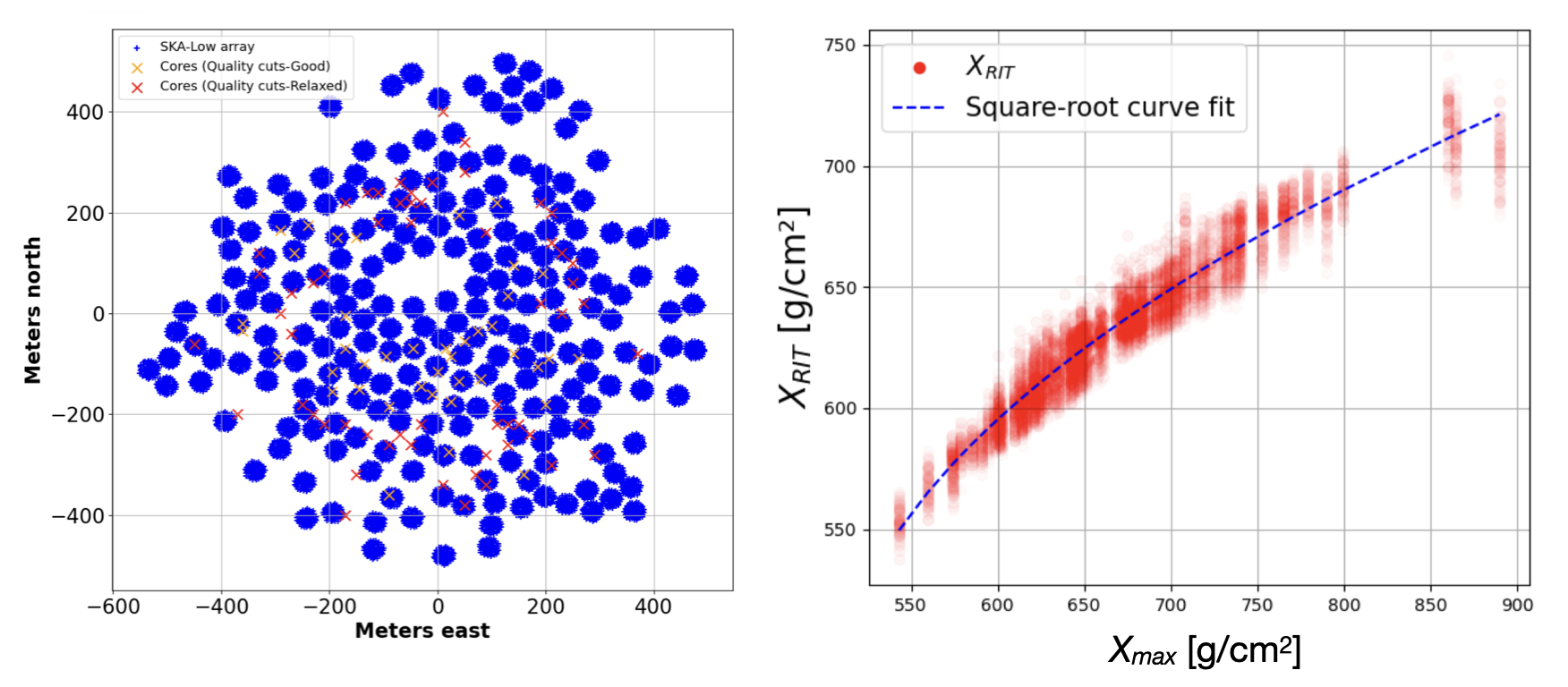}  
    \caption{Interferometric reconstruction of air showers at SKA-Low. Left: simulated shower core positions within the SKA-Low core. Right: Interferometric reconstruction of \Xrit, correlating with \Xmax.}
    \label{fig:SKA_RIT_reco}
\end{figure}

To demonstrate the potential for interferometric air shower reconstruction at SKA-Low we have performed a simulation study of 100 PeV air showers with 15$^{\circ}$ zenith angle arriving from the north, and with core locations throughout the core region of SKA-Low. The radio signal was generated with the CoREAS simulation (\cite{CoREAS:2013}), and the SKALA4 \citep{de_Lera_Acedo_2015} antenna response was applied to the simulated electric fields. Realistic noise contributions from the diffuse Galactic background (PyGSM) \citep{10.1111/j.1365-2966.2008.13376.x, 10.1093/mnras/stw2525} and the Rayleigh hardware noise (T $\sim$ 200 K) were injected. For each simulation, antennas with a position within 1.5 cherenkov radii have been included, and a selection has been made to include events with an even distribution of antennas around the shower core. Figure~\ref{fig:SKA_RIT_reco} shows the results of the RIT-based air reconstructions. The left panel shows the locations of the cores of the simulated shows with respect to SKA-Low AA*. The right panel shows the \Xrit reconstructions and their relation to the \Xmax values of the corresponding air showers. We see a strong correlation between \Xrit and \Xmax, highlighting the ability to reconstruct air shower development with RIT methods at SKA-Low.

\section{Summary}
Models of Galactic cosmic ray sources predict different mass composition fractions found at Earth, as a function of primary energy.
Hence, measuring this mass composition across the energy range of $10^{16}$ to $\unit[10^{18}]{eV}$ is an important science goal of a cosmic-ray observatory investigating the most energetic particle sources in the Galaxy. 
Building on experience and results from a decade of observations at LOFAR, we have shown that SKA-Low will be well suited to substantially improve these mass composition results. This is achieved by pushing the accuracy of individual shower measurements beyond current limits, by extending the energy range downward for full range coverage, and by accumulating sufficient statistics.
As we intend to use mainly the inner core region, the AA* array is sufficient to pursue all science goals.

The presented results serve as a starting point based on the techniques used at LOFAR.
Therefore, they put a lower bound on what can be achieved, and notable further improvements are expected.
The antenna density, two orders of magnitude higher than at LOFAR, allows for measuring cosmic-ray air showers to a level of detail that is unique in the field.
This helps for example to infer the longitudinal evolution of the air showers rather than just their maximum, allowing to overcome previous limitations to measuring the hydrogen/helium ratios more accurately.

So far, the analysis has been based on pulse energy measurements in individual antennas.
New analysis techniques are being developed to harness the full information in the radio signals. This allows for instance to take into account the information contained in the frequency spectra of the pulses, thus exploiting the wide frequency range of SKA-Low compared to LOFAR and other radio-based cosmic ray observatories. 
Directions being pursued include interferometry, being a natural technique for a dense radio array aiming to detect weak signals. And information field theory, a framework to infer physics results accurately from indirect measurements \citep{SKA_Book_ift}.

We conclude that the prospects for SKA-Low as a cosmic-ray observatory in parallel to normal astronomy operations are exciting, and at points described above, will reach beyond the current state of the art.

\section*{Acknowledgements}
The authors build on countless efforts to enable air shower observations using radio emission and are indebted to the community. Concretely, we acknowledge the following support: 
SBo, AN, and KT acknowledge the Verbundforschung of the German Ministry for Research, Technology and Space (BMFTR). 
PL and KW are supported by the Deutsche Forschungsgemeinschaft (DFG, German Research Foundation) – Projektnummer 531213488.
BH, CS, and PT are supported by ERC Grant Agreement No.~101041097.
KM acknowledges funding from the Netherlands Research School for Astronomy (NOVA) and Dutch Research Council (NWO) project OCENW.XS25.1.237. This research is supported by the Flemish Foundation for Scientific Research (FWO-AL991 and FWO-OZR4291).
ST acknowledges funding from the Khalifa University RIG-S-2023-070 grant.
The authors gratefully acknowledge the computing time provided on the high-performance computer HoreKa by the National High-Performance Computing Center at KIT (NHR@KIT). This center is jointly supported by the Federal Ministry of Education and Research and the Ministry of Science, Research and the Arts of Baden-Württemberg, as part of the National High-Performance Computing (NHR) joint funding program. HoreKa is partly funded by the German Research Foundation.

\bibliographystyle{abbrvnat-maxbibnames4}
\bibliography{chapter}

\end{document}